**Kinetic signatures of the region surrounding the X-line in asymmetric (magnetopause) reconnection**

M. A. Shay[1], T. D. Phan[2], C. C. Haggerty[1], M. Fujimoto[3], J. F. Drake[4], K. Malakit[5], P. A. Cassak[6], and M. Swisdak[4]

[1]Bartol Research Institute, Department of Physics and Astronomy, University of Delaware, Newark, Delaware, USA
[2]University of California, Berkeley, California, USA
[3]ISAS, Kanagawa, Japan
[4]University of Maryland, College Park, Maryland, USA
[5]Department of Physics, Mahidol University, Bangkok, Thailand
[6]Department of Physics, West Virginia University, Morgantown, West Virginia 26506, USA

**Abstract:** Kinetic particle-in-cell simulations are used to identify signatures of the electron diffusion region (EDR) and its surroundings during asymmetric magnetic reconnection. A "shoulder" in the sunward pointing normal electric field ($E_N > 0$) at the reconnection magnetic field reversal is a good indicator of the EDR, and is caused by magnetosheath electron meandering orbits in the vicinity of the x-line. Earthward of the X-line, electrons accelerated by $E_N$ form strong currents and crescent-shaped distribution functions in the plane perpendicular to B. Just downstream of the X-line, parallel electric fields create field-aligned crescent electron distribution functions. In the immediate upstream magnetosheath, magnetic field strength, plasma density, and perpendicular electron temperatures are lower than the asymptotic state. In the magnetosphere inflow region, magnetosheath ions intrude resulting in an Earthward pointing electric field and parallel heating of magnetospheric particles. Many of the above properties persist with a guide field of at least unity.

## 1. Introduction

Magnetic reconnection occurs in a small diffusion region around the X-line but its consequences are large scale. Understanding kinetic processes in the diffusion regions for both symmetric (magnetotail-like) and asymmetric (magnetopause-like) reconnection is the primary objective of the current Magnetospheric Multiscale (MMS) mission. In the context of this paper, we define the electron diffusion region (EDR) to be the electron-scale region surrounding the x-line in which magnetic connectivity is ultimately broken. Note that this definition is fundamentally non-local in nature.

Diffusion regions are minuscule so their detections by spacecraft are rare. A challenge for spacecraft observations is the identification of the diffusion region in data. There are different approaches to determine whether or not a spacecraft has encountered the diffusion region. One approach is to identify the diffusion region based on theoretically expected kinetic signatures of diffusion region: enhanced dissipation [e.g., Zenitani et al., 2012]; non-gyrotropic particle behavior [e.g., Scudder et al., 2008; Aunai et al., 2013; Swisdak 2016]; or electron distribution functions [e.g., Chen et al., 2008, Ng et al., 2011]. While this approach is required to reveal diffusion region processes, some of the kinetic signatures are difficult to measure accurately in experiments. Furthermore, some signatures also exist downstream of the diffusion region and along the separatrices.

An alternative and complementary approach, taken in the present paper, is to identify diffusion region candidates by a combination of large-scale context, e.g., the properties of the region surrounding the ion and electron diffusion regions, and simple to measure signatures of the electron diffusion region itself. Such a scheme provides crucial consistency checks for the interpretation of diffusion region encounters based on observed kinetic signatures.

This paper addresses ways to recognize proximity to the diffusion region at the magnetopause where reconnection involves asymmetric inflow conditions. We focus on the case of anti-parallel reconnection (i.e., no guide field) but many of the signatures described are also present in guide field reconnection.

## 2. Simulations

We perform a particle-in-Cell (PIC) simulation [Zeiler et al., 2002] of asymmetric magnetic reconnection with no guide field. Magnetic field strengths and particle number densities are normalized to $B_0$ and $n_0$, respectively. Lengths are normalized to the ion inertial length $d_{i0} = c/\omega_{pi0}$ at the reference density $n_0$, time to the ion cyclotron time $(\Omega_{ci0})^{-1} = m_i c / (e B_0)$, and velocities to the Alfven speed $c_{A0} = d_{i0} / \Omega_{ci0}$. Electric fields and temperatures are normalized to $E_0 = c_{A0} B_0 / c$ and $T_0 = m_i c_{A0}^2$, respectively.

The simulations are 2 ½ dimensional and performed in the L ⨯ N plane of the LMN current sheet coordinate system, with the L being the reconnection outflow direction, N being the inflow direction, and M along the X-line direction. The simulation domain size is 102.4 $d_i$ × 51.2 $d_i$ with 2048 × 1024 grid cells, 50 particles per grid cell, $c/c_{A0} = 15$, and $m_i/m_e=25$. The initial conditions are an asymmetric double current sheet (Malakit et al., 2010). The simulation uses magnetosheath to magnetosphere density and magnetic field ratios of 10 and ½ respectively. $T_i/T_e = 2$ with $T_i = 1.33$ in the magnetosheath and 3.33 in the magnetosphere. A small magnetic perturbation is used to initiate reconnection. The simulation is evolved until reconnection reaches a steady state, and then for analysis purposes during this steady period the simulation data is time averaged over one ion cyclotron time $(\Omega_{ci0})^{-1}$.

### 3. Electron diffusion region signatures: Normal electric field shoulder

Figure 1 shows simulation results in the L-N plane: magnetic and electric fields, electron and ion flows, current, density, temperatures, and different measures of the violation of the ion and electron frozen-in conditions. The solid contours are magnetic field lines and the dotted line is the midplane where $B_L = 0$. Electron diffusion regions should exhibit a number of properties such as (1) the violation of the electron frozen-in condition, (2) non-gyrotropic electron distributions [e.g., Scudder et al., 2008; Aunai et al., 2013; Swisdak 2016], and (3) enhanced dissipation [Zenitani et al., 2011]. However, the violation of the frozen-in condition and non-gyrotropic distributions in themselves do not uniquely define the electron diffusion region at the X-line. In Figure 1m-p are shown 2D plots of $\mathbf{E} + \mathbf{V} \times \mathbf{B}$ for both the ions and electrons, along the N and M directions. In the N direction (Figure 1m-n), both the electrons and ions show large values near the magnetospheric separatrices [Mozer et al., 2009]. In the M direction (Figure 1o-p), although $\mathbf{E} + \mathbf{V} \times \mathbf{B}$ peaks close to the x-line, it has significant value throughout the exhaust, even for the electrons.

A measure of non-gyrotropy $D_{ng}$ [Aunai et al., 2013, Eq. 2] as shown in Figure 1q, is also not localized close to the x-line. There are significant regions of non-gyrotropy along both the magnetospheric and the magnetosheath separatrices. A frame independent dissipation measure $D_e = \mathbf{J} \cdot (\mathbf{E} + \mathbf{V}_e \times \mathbf{B}) - (n_i - n_e)(\mathbf{V}_e \cdot \mathbf{E})$ [Zenitani et al., 2011] is strongly peaked very close to the x-line in Figure 1r and is therefore a good parameter with which to identify the electron diffusion region, although $D_e$ is slightly enhanced along the separatrices as well. Figures 1s and t show the non-negligible terms of $D_e$: $E_\parallel J_\parallel$ and $\mathbf{J}_\perp \cdot (\mathbf{E}_\perp + \mathbf{V}_e \times \mathbf{B})$. Care must be taken when decomposing $\mathbf{J} \cdot \mathbf{E}$ into perpendicular and parallel terms, as in this case the large positive $E_\parallel J_\parallel$ near the X-line is mostly canceled by the negative perpendicular term such that $\mathbf{J} \cdot \mathbf{E}$ is nearly zero in this region.

Our simulation reveals a simple to measure indicator of the electron diffusion in the form of a region of overlap between the sunward pointing normal electric field ($E_N > 0$) and the field reversal region $B_L = 0$. Below we denote this as the "overlap" or "shoulder" region. Note that $E_N$ is the largest component of the electric field at the magnetopause [Vaivads et al., 2004, Pritchett, 2008, Tanaka et al., 2008]. Figure 2a-j show 1-D profiles of various quantities along N through the x-line. The strong $E_N$ has a visible shoulder and is >0 at the midplane, which is the x-line in this cut.

While $E_N > 0$ occurs all along the magnetospheric separatrix, extending long distances downstream of the X-line, $E_N > 0$ only touches the mid-plane close to the X-line (within 2.75 $d_i$ along the outflow direction) and it exhibits a shoulder that is not seen away from the EDR. In Figure 3b-c are shown cuts of $E_N$ at 1.75 $d_i$ and 2.75 $d_i$ ( 8.75 and 13.75 $d_e$ ) downstream of the x-line. $E_N$ is now displaced toward the magnetosphere side of the reversal region and by 2.75 $d_i$ downstream no longer overlaps with $B_L = 0$. From Figure 1c,q and r, the overlap/shoulder region coincides with the region of enhanced dissipation measure $D_e$ and where the electrons are non-gyrotropic, suggesting that this is a good indicator of the electron diffusion region. Indeed, it will be shown in the following paragraphs that the physics ultimately creating the shoulder is intimately linked to the electron kinetic physics associated with the electron diffusion region.

We now examine the kinetic behavior of the electrons in the region in the vicinity of the X-line. The stagnation point of the electron flow occurs on the magnetospheric side of the x-line [Cassak et al., 2007] and marks the transition between magnetosheath and magnetosphere plasma. Here there are strong gradients in density and electron temperature, as well as a peak in $V_{eM}$ (Figure 2c-e). Examination of the N component of Ohm's law reveals that the physics of the overlap/shoulder region is fundamentally different from the rest of the strong $E_N$ region (that extends along the separatrices). The typical strong $E_N$ is characterized by a large $(\mathbf{V}_e \times \mathbf{B})_N$ which is partially offset by a $(\nabla \bullet \mathbf{P}_e)_N$ of the opposite sign in Figures 2h and 3q. In the shoulder region in Figure 2h, however, $(\mathbf{V}_e \times \mathbf{B})_N$ becomes small and the electron pressure term changes sign, creating the shoulder on $E_N$ at the x-line.

The electron pressure gradient along N in this region is due to variation in $P_{eNN}$ (Figure 2i). Approaching from the magnetospheric side, the magnetic pressure drops precipitously and is offset primarily by the increased pressure of the magnetosheath plasma. The same behavior is seen far downstream of the x-line. However, in the shoulder region, the electron pressure gradient is also produced by a gradient of $T_{eNN}$, as shown in Figure 2j; quite striking also is that the peak in $T_{eNN}$ is straddled by two peaks of $T_{eMM}$.

This structuring of the electron diagonal pressure terms is due to the magnetosheath electron orbits associated with the sharp gradients in the EDR. In Figure 2p is a schematic in the M-N plane of the magnetosheath electron motion in the vicinity of the x-line. Sheath electrons cross the x-line, are accelerated by $E_N$ and then turned by the magnetosphere $B_L$ into the M direction, and then return to the x-line. The motion is very similar to the cusp-like motion of pickup ions in the solar wind. This cusp-like motion in $E_N$ rather than the usual meandering motion at the X-line is responsible for the crescent velocity distribution (Figures 2l-n) seen in previous studies [Hesse et al., 2014]. Close to the x-line in Figures 2m-n, a full crescent shape is created by sheath electrons flowing both towards and away from the x-line along N, creating a peak in $T_{eNN}$ in Figure 2j. Deeper into the magnetosphere (Figure 2l), electrons are accelerated to higher energy creating a crescent at higher velocities. The electrons counterstreaming along M relative to the magnetospheric population creates a strong peak in $T_{eMM}$. Note also that these crescent distribution functions straddle the location of the peak of $V_{eM}$ and $J_m$. This strong current causes a large change in $B_L$ necessary to balance the large pressure gradient in this region.

Sheath electrons that have reflected from the magnetosphere side cross the X-line onto magnetosheath field lines (Figure 2p) and are accelerated in the M direction by the reconnection electric field $E_M$. The motion of these high velocity sheath electrons relative to the newly incoming sheath electrons creates counter-streaming electron beams along the M direction in the shoulder region, as see in Figure 2o, and leads to a second peak of $T_{eMM}$ associated with the shoulder (point 4 in Fig. 2j). These counterstreaming beams are created by the proximity of the $B_L = 0$ region to the strong $E_N$ region (the shoulder), which creates the electron meandering motion [Horiuchi and Sato, 1994] unique to the electron diffusion region very close to the x-line. This shoulder of EN, because it is associated with kinetic electron orbits, has a width (along the N direction) of around 2 $d_e$, and is expected to have a width comparable to electron scales when observed in the magnetosphere.

Unlike the shoulder of $E_N$ and the associated counter-streaming electron beams along M, the crescent-shaped electron distribution functions are not nearly as localized around the X-line. Shown in Figure 3d-i are distribution functions and 1D profiles at 6.35 $d_i$ downstream of the x-line. As expected there is no overlap between $E_N>0$ and $B_L=0$ (Figure 3l), no secondary peak of $T_{eMM}$ (Figure 3r), and no counter-streaming electron beams associated with meandering orbits (Figure 3h). However, the strong peak in $T_{eMM}$ on the magnetosphere side of the reversal region and the gradient of $T_{eNN}$ associated with the large $E_N$ are still present in Figure 3r. The phase space density plots still reveal the crescent-shaped distribution functions, with the crescent becoming almost circular but nonuniform in Figure 3f. Associated with this is a "parallel outflow crescent" in ($V_\parallel$, $V_{\perp 2}$) space (Figure 3g). Notably, the peak velocity of this outflow crescent is the same as the crescent velocity around the x-line. The streaming velocity results from a weak parallel electric field, which arises from the small value of $B_N$ within the region of high $E_N$. The band of strong $E_N$ downstream lies inside (magnetosheath side) of the separatrix so the total potential drop along B downstream is the same as the potential drop along N at the x-line. The parallel electron streaming velocity therefore matches the peak $V_M$ at the x-line. This outflow crescent starts to exist in the $E_N$ shoulder only $1\ d_i = 5\ d_e$ downstream of the X-line (not shown) and extends relatively far downstream (Figure 3e,g). Having spacecraft observations of oppositely directed "parallel outflow crescents" would be evidence of straddling the X-line. MMS has observed both crescent and full-circle distributions, and well as oppositely directed parallel outflow crescents [Burch et al., 2016] which would place the spacecraft in close proximity of the X-line.

## 4. Theoretical Model for Crescent Shaped Phase Space Density

This crescent distribution in Figures 2l-n results from cusp-like orbits of electrons associated with the motion in the (M,N) plane controlled by $E_N(N)$ and $B_L(N)$ in the magnetosphere. In Figure 4a is a schematic showing both the simplified form of $E_N$ and $B_L$ and the electron particle motion. $E_N$ and $B_L$ are assumed to be zero on the magnetosheath side of the x-line and then increase linearly with distance N into the magnetosphere.

We focus on the motion in the (M, N) plane with a simple model in which $E_N = E'_N N$ and $B_L = B'_L N$ with the prime denoting a derivative with respect to N. The equations of motion can be integrated once to obtain two constants of motion, the energy $W_N (V_N, N)$ of electrons moving in the N direction and the canonical momentum $P_M (V_M, N)$ in the M direction,

$$W_N = V_N^2 - 2(V_{EB} - V_{M0}) V_{EB} \frac{N^2}{N_0^2} + V_{EB}^2 \frac{N^4}{N_0^4} = V_{N0}^2 \tag{1}$$

$$P_M = V_M - V_{EB} \frac{N^2}{N_0^2} = V_{M0}, \tag{2}$$

where $V_{EB} = cE'_N/B'_L$ is the E×B velocity in the M direction and $N_0 = \sqrt{-2 V_{EB} m_e c/(eB'_L)}$ is the characteristic spatial scale length of the cusp-like orbits. Note that $V_{EB}$ is positive while $B'_L$ and $E'_N$ are negative. $V_{N0}$ and $V_{M0}$ are the initial velocities of the electrons at N = 0. The constancy of $W_N$ implies that as a function of N, $V_N$ first increases and then eventually goes to zero as the electrons are accelerated by $E_N$ and then turned by $B_L$ to return to N = 0. $V_M$ reaches its maximum value at the maximum excursion of the electron in the N direction.

The electron distribution function $f(V_M, V_N)$ in the region of finite $E_N$ can be written in terms of the constants of motion $W_N$ and $P_M$. We choose a form that produces a Maxwellian distribution of initial velocities $V_{N0}$ and $V_{M0}$:

$$f(V_M, V_N) \propto e^{-P_M^2/v_{the}^2} e^{-W_N/v_{the}^2} \left(1 + \tanh \frac{W_N^2}{v_{the}^2}\right)/2, \tag{3}$$

where $v_{the}$ is the electron thermal velocity at $N = 0$, $V_{M0}$ in Eq. (2) is used in $W_N$ and the tanh function forces $W_N$ to be positive. The only free parameter is $V_{EB}/v_{the}$. For the simulations shown in Fig. 2, $V_{EB} \approx 3.0$ and $N_0 \approx 0.4$. In Fig. 2q-s we show three modeled distribution functions at $N/N_0 = 0.75$, 1.0, and 1.5. The distributions are symmetric in $V_N$, corresponding to particles moving toward higher N and then returning. With increasing N there is a transition from a Maxwellian to a horseshoe distribution to a crescent with a peak at an increasing value of $V_M$ but a narrower width along $V_M$.

## 5. Signatures of the large-scale context surrounding the X-line

The two inflow regions close to the X-line exhibit distinct properties that can help identify the proximity to the X-line and provide the context for satellite observations.

### 5.1. High-density (magnetosheath) inflow region

In typical magnetopause reconnection, the much weaker magnetic field of the inflowing magnetosheath plasma has a much larger inflow velocity ($V_{iN}$) compared with that on the magnetospheric side. This large inflow velocity leads to substantial bending of the magnetosheath field upstream of the x-line (Figure 1), and results in an increased magnetic tension force away from the x-line which leads to a reduction in total pressure relative to ambient magnetosheath conditions; both the reconnection magnetic field $B_L$ and the plasma density $n_i$ are decreased in this region of curved field lines. The spreading of the magnetic field also leads to a reduction in $T_{e\perp}$ (Figure 1l) and $T_{i\perp}$ (not shown).

### 5.2. Low-density (magnetospheric) inflow region

An Earthward pointing "Larmor" electric field (cyan region in Figure 4b) exists in the magnetosphere inflow [Malakit et al., 2013] as well as enhanced $T_{e\|}$ (Figure 4c) [Egedal et al., 2011]; the electric field is responsible for generating the $T_{e\|}$ and both exist within ~ 15 $d_i$ of the x-line [Malakit et al., 2012]. We have explored other signatures in this region and found that the region is also characterized by the intrusion of the ion outflow ($V_{iL}$) jet (Figure 1d) and the out-of-plane $V_{iM}$ (Figure 4c), while the electron flows (Figures 1f and g) are mostly confined to the magnetosheath side of the separatrix.

The distinct ion and electron flow behavior suggests that the penetration of the magnetosheath ions into the magnetosphere is due to an overshoot of the bulk motion of magnetosheath ions into the magnetosphere. This overshoot occurs because for asymmetric reconnection with large density asymmetry, the stagnation point is located on the magnetospheric side of the reconnection layer [Cassak and Shay, 2007]. This overshoot from the finite ion Larmor gyroradius results in a net out-of-plane drift (i.e., $V_M$) of the magnetosheath ions as can be seen in the ion distributions (Figure 4e-h). Associated with this negative $V_M$ flow (and the positive $B_L$) is an earthward pointing normal electric field.

The lack of magnetosheath electron penetration into the magnetosphere is due to their smaller gyroradii. To achieve charge neutrality in the Larmor region where there is an excess of magnetosheath ions, the magnetospheric electrons are accelerated into the region by a parallel electric field, which results in a large parallel electron temperature increase in the region (Figure 4d). A key evidence for the magnetospheric electrons being responsible for the parallel temperature increase is shown in electron energy flux in Figures 4i-l. The electron core energy is about 3.1 in the magnetosphere proper (Figure 4i) and 1.6 in the magnetosheath proper (Figure 4l). The electron distributions in the Larmor electric field region display strong field-aligned anisotropy and have a peak electron core energy of 3.85 in Figure 4k, higher than the core energy of magnetospheric electrons which is consistent with them being heated magnetospheric electrons rather than heated magnetosheath electrons.

From the 1-D profiles in Figure 2a-l, a spacecraft crossing normal to the magnetopause would detect no outflow ($V_L \sim 0$) and no out-of-plane Hall magnetic field, but would detect out-of-plane ion flow ($V_M$) in the magnetospheric inflow region, the associated Larmor ($E_n < 0$) electric field, and enhanced $T_{e\parallel}$. From Fig. 3j-r 6.35 $d_i$ downstream of the x-line, a spacecraft crossing this region would still measure the Larmor electric field and other associated signatures, but would also see a clear intrusion of $V_{iL}$ into the magnetospheric inflow region (upstream the strong $E_N$).

### 6. Summary and Discussions:

Using 2-D PIC simulations, we have examined signatures of the asymmetric reconnection electron diffusion region and its surroundings. We emphasize signatures that are relatively easy to measure experimentally. A simple and practical indicator of the electron diffusion region is the presence of a sunward pointing $E_N$ at the midplane (called the "shoulder") as this signature coincides with the region of enhanced dissipation, non-gyrotropic electrons at midplane, and counterstreaming electron beams due to electron meandering orbits around the X-line. This $E_N$ signature is straightforward to measure experimentally because it is the largest component of the electric field at the magnetopause.

Crescent shaped electron distribution functions in ($V_{\perp 1}$, $V_{\perp 2}$) plane and "parallel outflow crescents" in ($V_\parallel$, $V_{\perp 2}$) plane are associated with the strong $E_N$ which occurs in asymmetric magnetic reconnection. While these signatures are not as localized as the $E_N$ shoulder, spacecraft straddling the x-line would observe oppositely directed "outflow crescents". The various types of crescent distributions have recently been observed by MMS in the vicinity of an X-line [Burch et al., 2016].

On a larger scale in the magnetospheric inflow region, a Larmor electric field is caused by the intrusion of magnetosheath ions into the magnetospheric inflow region and the resulting net out-of-plane flows of the penetrating magnetosheath ions. To preserve charge neutrality, magnetospheric electrons are drawn into the region, resulting in the enhancement of electron parallel temperature and an associated temperature anisotropy. The detection of the plasma and field signatures associated with the Larmor effect, including field-aligned temperature anisotropy, would imply that the spacecraft is within 15 ion skin depths of the X-line.

On the magnetosheath side of the inflow region, the magnetic field magnitude, plasma density, and electron temperature are reduced compared to their upstream (asymptotic) values.

### Acknowledgments

This research was supported by NSF Grants AGS-1202330 and AGS-0953463; NASA Grants NNX08A083G–MMS IDS, NNX14AC78G, NNX13AD72G, and NNX15AW58G; and the UDel NASA Space Grant. Simulations and analysis were performed at NCAR-CISL and at NERSC. The data used are listed in the text, references, and are available by request.

**Figure Captions**

Figure 1: Simulation results in the LN plane. Quantities plotted denoted as text in each frame. $D_{ng}$ is a non-gyrotropy measure [Aunai et al., 2013] and $D_e$ is a dissipation measure [Zenitani et al., 2012. Solid black contours are magnetic field lines and dotted lines show the midplane (defined where $B_L = 0$).

Figure 2: (Left Column): 1-D spatial profiles along N, on the dashed line through the x-line shown in the 2-D image in (k). (k) $E_N$, vertical dashed line is location of 1-D cuts, dotted line is midplane, rectangles denote locations of distribution functions. (l-o) Electron distribution functions in $(V_M, V_N)$ plane. Distributions are integrated between $V_e = \pm 3$ (in electron bulk flow frame) along third velocity direction $V_L$. Spatial domain sampled to create distributions shown in title of each panel. (p) Schematic of magnetosheath electron motion in vicinity of X-line. (q-s) Distribution functions predicted from electron motion in linear ramp model of $E_N$ and $B_L$. In (p), the corresponding positions of the phase space densities (l-o) and temperatures (j) are denoted by the circled numbers 1-4.

Figure 3: (a) $E_N$, vertical dashed line is location of 1-D cuts, dotted line is midplane, rectangles denote locations of distribution functions, (b-c) 1-D spatial profiles of E and $J_M$ at 1.75 and 2.75 $d_i$ downstream of x-line. (d)-(i) Electron distribution functions in $(v_{\perp 1}, v_{\perp 2})$ and $(v_\parallel, v_{\perp 2})$ planes, integrated between $v = \pm 3$ along third velocity direction (in electron bulk flow frame); $\mathbf{v}_{\perp 1}$ along $\mathbf{E} \times \mathbf{B}$ direction and $\mathbf{v}_{\perp 2}$ along $\mathbf{B} \times (\mathbf{E} \times \mathbf{B})$. (j-r) 1-D spatial profiles along N taken 6.35 $d_i$ downstream of x-line. Vertical dashed line is location of midplane.

Figure 4: (a) Schematic of electron motion in linear ramp model with spatial variation of $B_N$ and $E_L$. (b-d) Spatial profiles of $E_N$, $V_{iM}$, and $T_{e\parallel}$. Dotted line is the midplane with boxes showing sampling locations used to create ion distribution functions and electron energy flux. (e-h) Ion distribution functions. Domain of sampling shown in title of each frame; integrated between $v = \pm 0.6$ along third direction (in ion bulk flow frame). (i-l) Electron energy flux dependence on energy and pitch angle $\theta$. Domain sampling same as ion distribution function above each energy flux panel. Vertical dashed line is electron core energy.

Khotyaintsev, M. Andre, M. B. Bavassano-Cattaneo, S. C. Buchert, and C. J. Owen (2008), Effects on magnetic reconnection of a density asymmetry across the current sheet, *Annales Geophysicae,* 26, 2471, doi:10.5194/angeo-26-2471-2008.
15. Vaivads, A., Y. Khotyaintsev, M. Andre, A. Retino, S. C. Buchert, B. N. Rogers, P. Decreau, G. Paschmann, and T. D. Phan (2004), Structure of the Magnetic Reconnection Diffusion Region from Four-Spacecraft Observations, *Physical Review Letters,* 93, 105001, doi:10.1103/PhysRevLett.93.105001.
16. Zeiler, A., D. Biskamp, J. F. Drake, B. N. Rogers, M. A. Shay, and M. Scholer (2002), Three-dimensional particle simulations of collisionless magnetic reconnection, *Journal of Geophysical Research (Space Physics),* 107, 1230, doi:10.1029/2001JA000287.
17. Zenitani, S., M. Hesse, A. Klimas, and M. Kuznetsova (2011), New Measure of the Dissipation Region in Collisionless Magnetic Reconnection, *Physical Review Letters,* 106, 195003, doi:10.1103/PhysRevLett.106.195003.

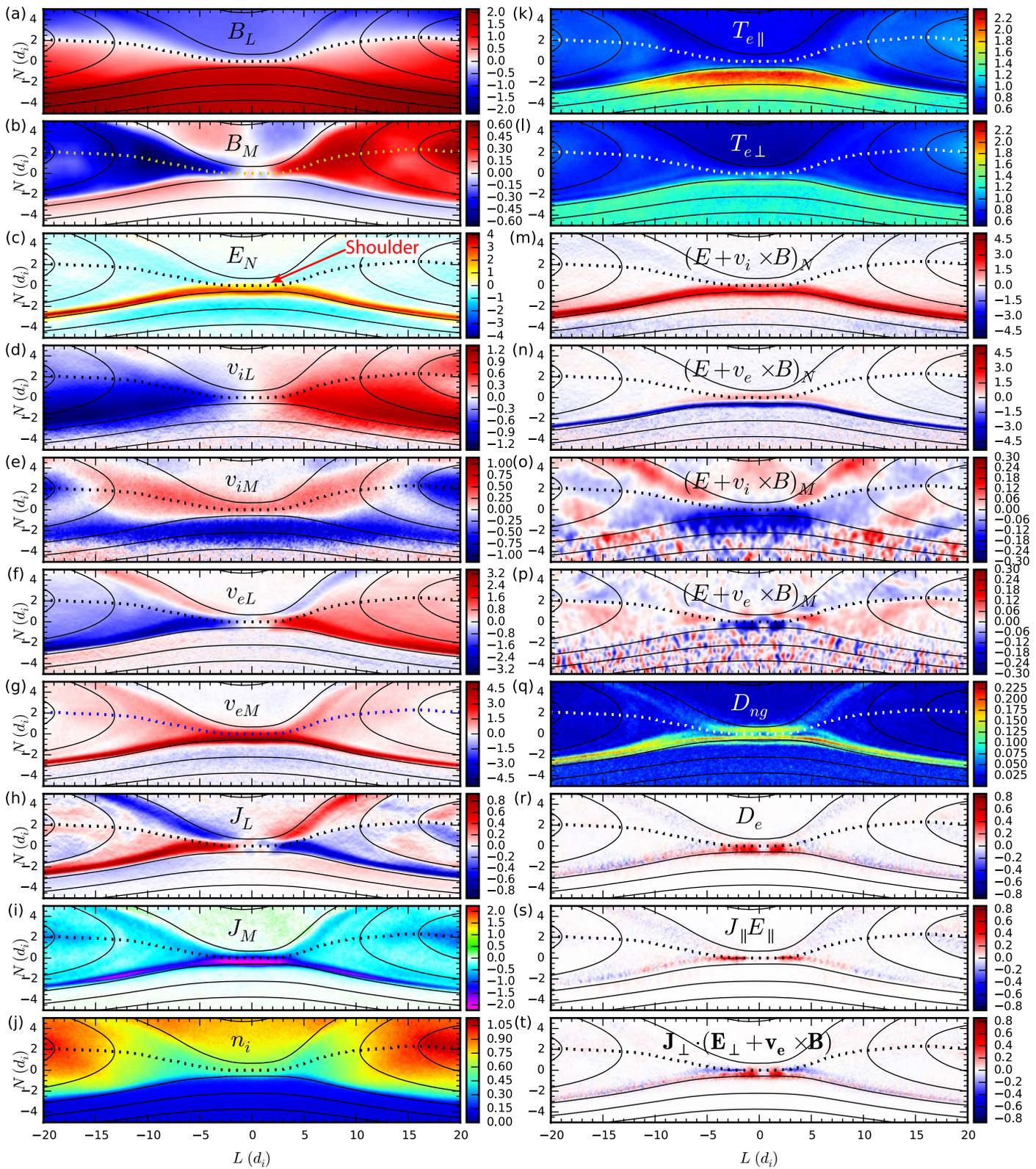

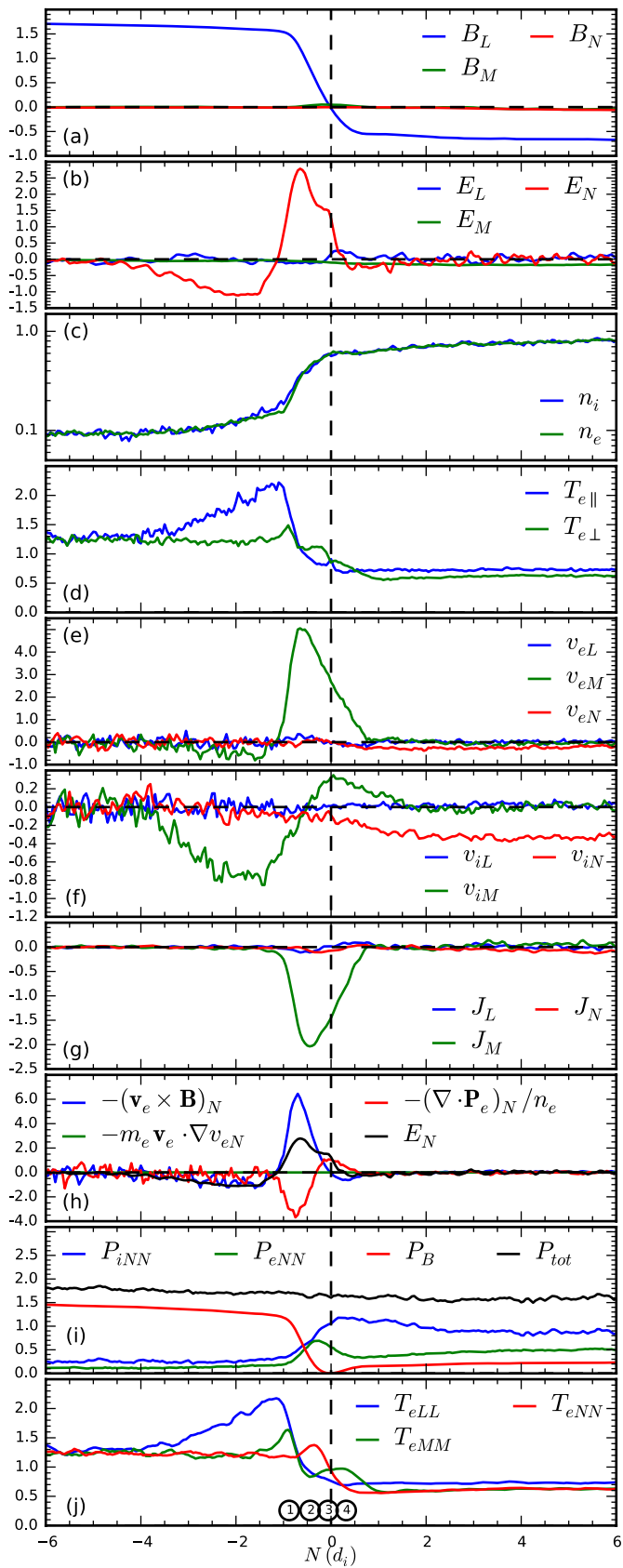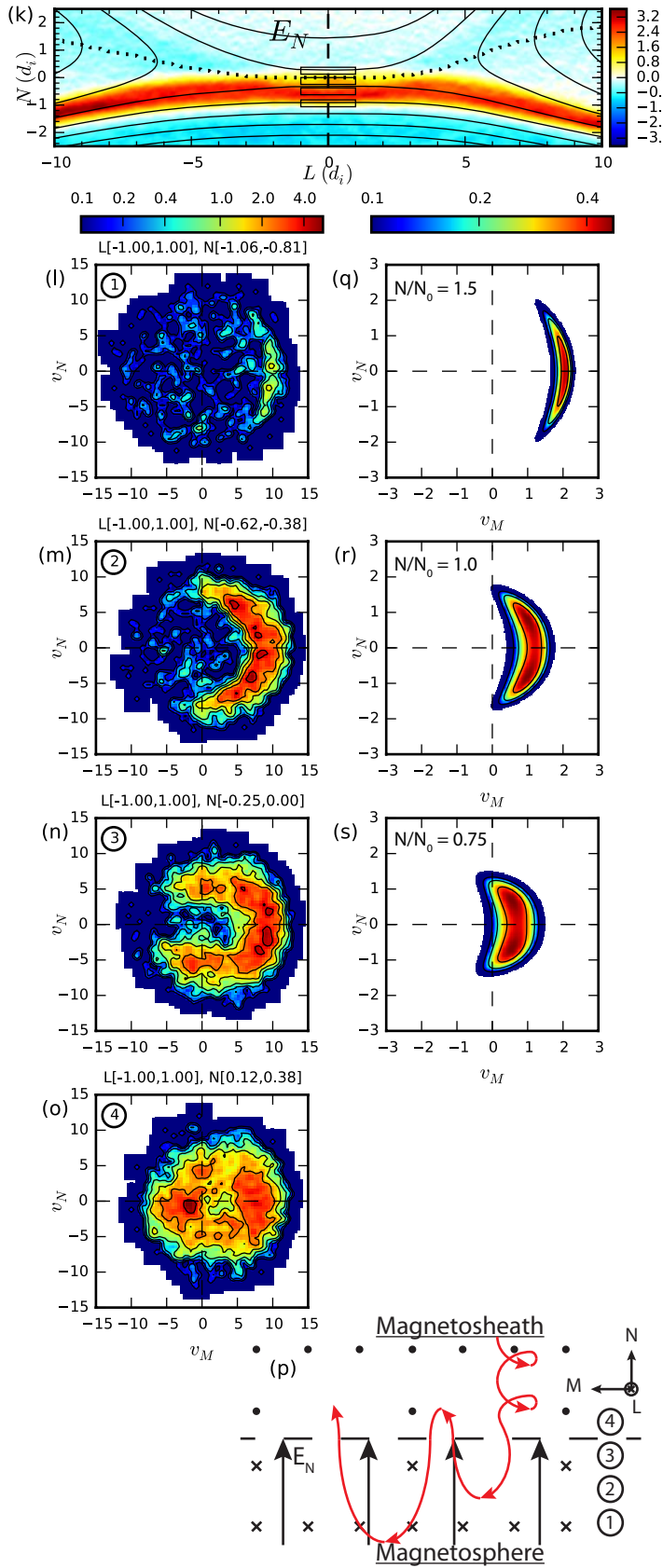

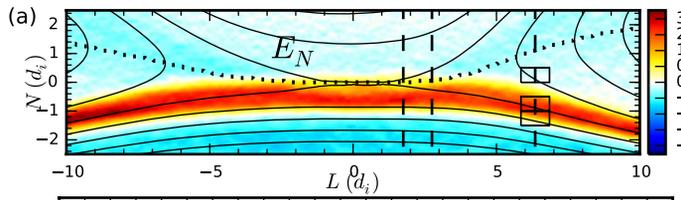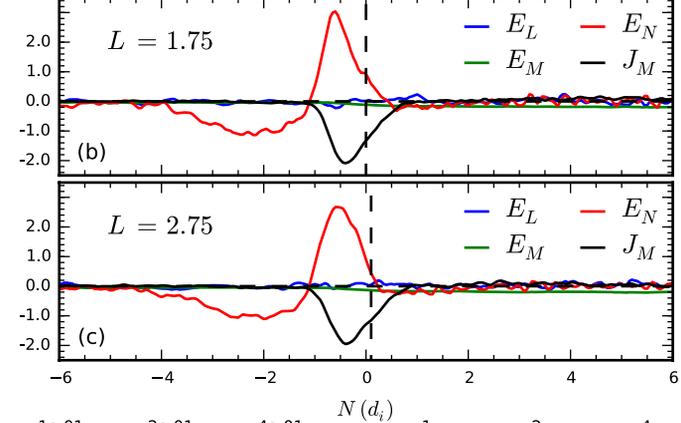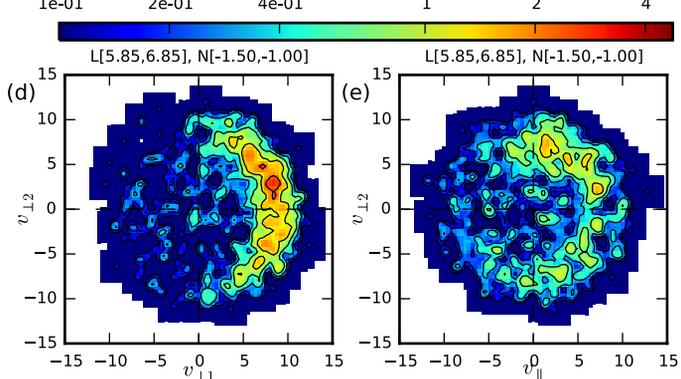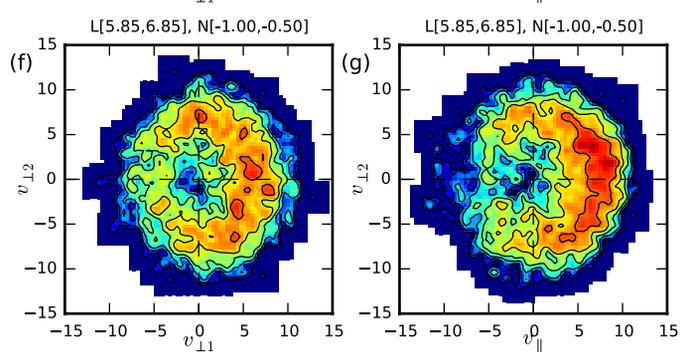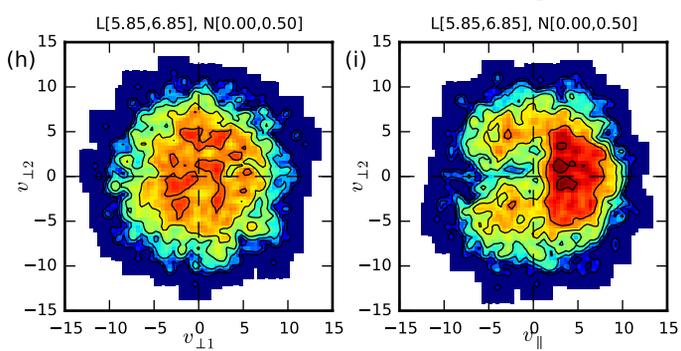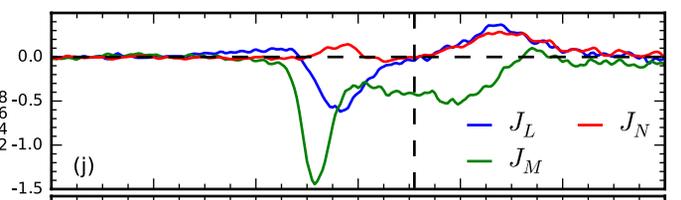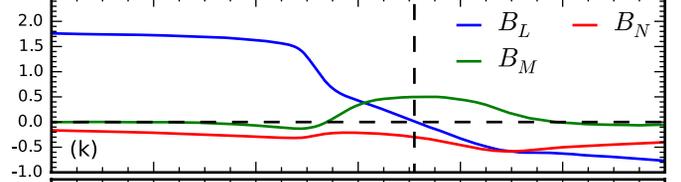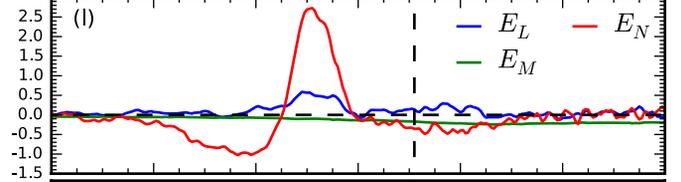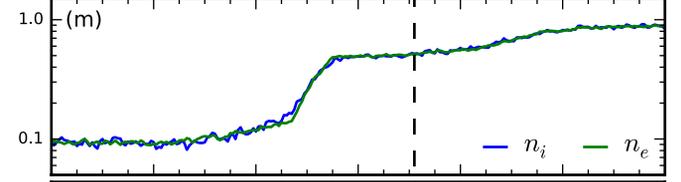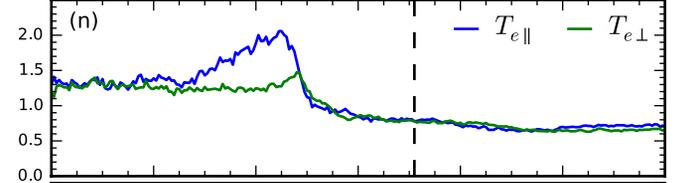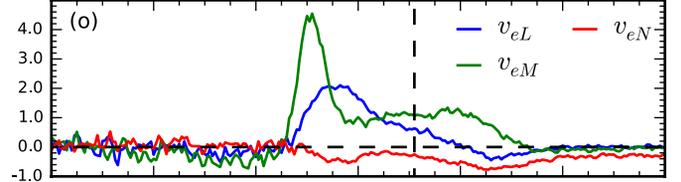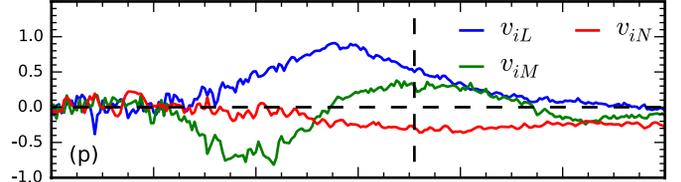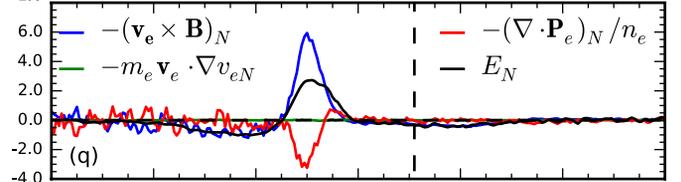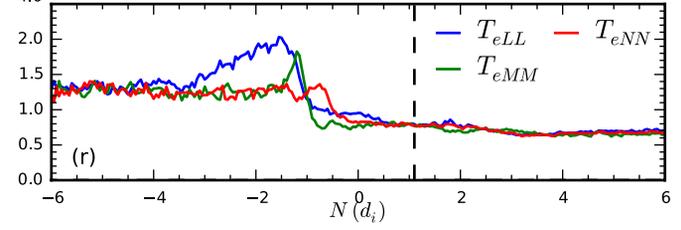

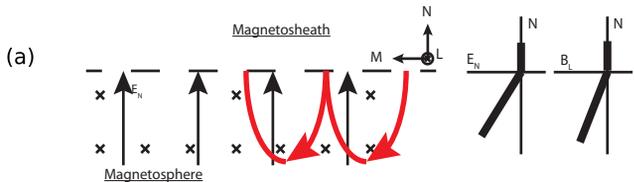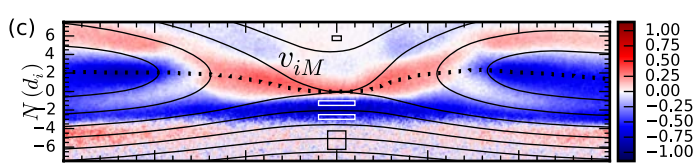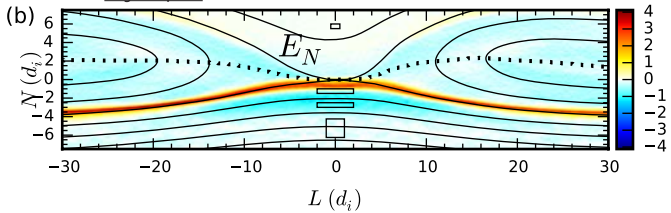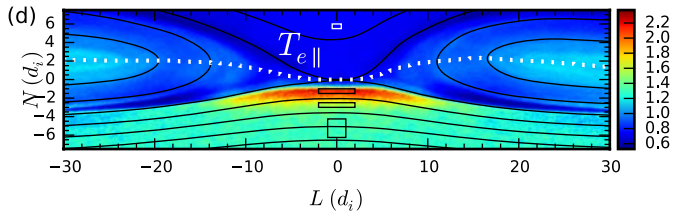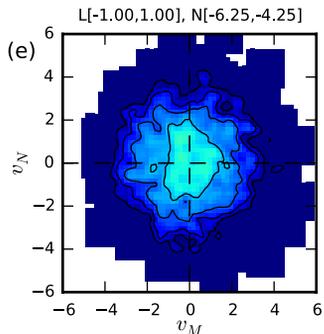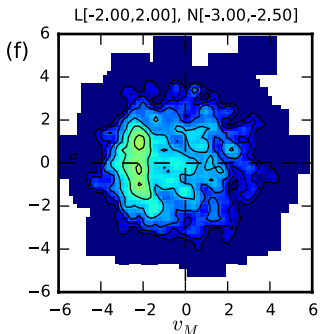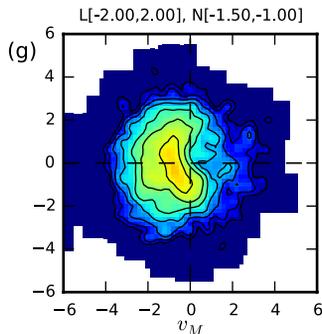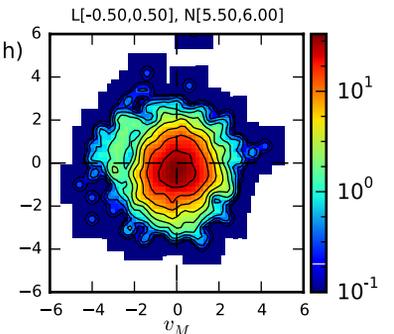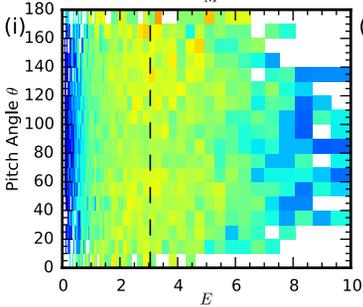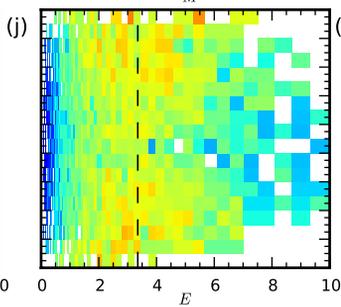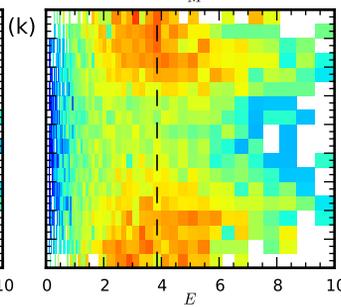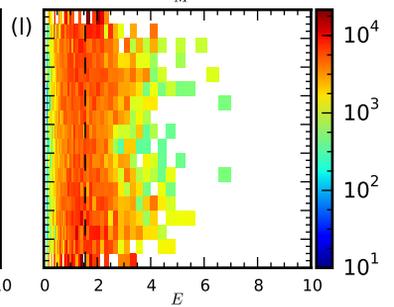